\documentclass[aps,prb,twocolumn,groupedaddress,showpacs]{revtex4-1}
\usepackage{epsfig}
\usepackage[dvipsnames,usenames]{color}
\usepackage[normalem]{ulem}
\usepackage{amsmath}
\usepackage{xfrac}
\emergencystretch=\maxdimen
\begin{document}

\title{Impurities near an Antiferromagnetic-Singlet
Quantum Critical Point}

\author{T. Mendes$^{1,2}$, 
N.~Costa$^1$, 
G.~Batrouni$^{3,4,5,6}$,
N.~Curro$^2$,
R.R.~dos Santos$^1$,
T.~Paiva$^{1}$, 
and R.T.~Scalettar$^{2}$}

\affiliation{$^1$Instituto de Fisica, Universidade Federal do Rio de
Janeiro Cx.P. 68.528, 21941-972 Rio de Janeiro RJ, Brazil}

\affiliation{$^2$Physics Department, University of California, Davis,
California 95616, USA}

\affiliation{$^3$
Universit\'e C$\hat o$te d'Azur, INLN, CNRS, France
}

\affiliation{$^4$
Institut Universitaire de France, 75005 Paris, France
}

\affiliation{$^5$
MajuLab, CNRS-UNS-NUS-NTU International Joint Research Unit UMI 3654,
Singapore
}

\affiliation{$^6$
Centre for Quantum Technologies, National University of Singapore; 2
Science Drive 3 Singapore 117542
}

\begin{abstract}
Heavy fermion systems, and other strongly correlated electron materials,
often exhibit a competition between antiferromagnetic (AF) and singlet
ground states.  Using exact Quantum Monte Carlo (QMC) simulations, we
examine the effect of impurities in the vicinity of such AF-singlet
quantum critical points, through an appropriately defined ``impurity
susceptibility," $\chi_{\rm imp}$.  
Our key finding is a connection, within a
single calculational framework, between AF domains induced on the
singlet side of the transition, and the behavior of the nuclear magnetic
resonance (NMR) relaxation rate $1/T_1$.  We show that local NMR
measurements provide a diagnostic for the location of the QCP which
agrees remarkably well with the vanishing of the AF order parameter
and large values of $\chi_{\rm imp}$.  We
connect our results with experiments on Cd-doped CeCoIn$_5$. 
\end{abstract}


\maketitle

\noindent

\underbar{Introduction:}
In materials like the cuprate superconductors, mobile impurities,
introduced {\it eg} via the replacement of La by Sr, are known to
destroy antiferromagnetic (AF) order very
rapidly\cite{Takagi89,Uchida91}.  Long range spin correlations are
somewhat more robust to static scatterers, {\it eg} via Zn substitution
for Cu in the same materials\cite{xiao88,keimer92,mahajan94}.  This
competition of AF and chemical doping is a central feature of many other
strongly correlated systems, including Li doping in nickel
oxides\cite{ido91,reinert95}, spin chains\cite{martin97}, and
ladders\cite{azuma97}, and has been explored by QMC approaches in
single band fermion models\cite{ulmke99} and their strong coupling spin
limits\cite{hoglund04}.

Materials with multiple fermionic bands or localized spins in
multi-chain or multi-layer geometries offer an additional richness to
the problem of the effect of impurities on AF, since even in the clean
limit they can exhibit a quantum critical point (QCP) separating AF and
singlet phases.  Disorder introduces a non-trivial effect on this
transition:  Although impurities reduce AF deep in the ordered phase,
nearer to the QCP they can increase AF and even, 
beginning in the quantum disordered
phase, {\it induce} AF by breaking
singlets\cite{martin97,azuma97}.

This effect has recently been explored in heavy-fermion materials
where, e.g.~Cd doping of superconducting CeCoIn$_5$ induces long range
magnetic order\cite{pham06}.   The underlying mechanism is believed to
be a local reduction of conduction electron-local moment hybridization
on the Cd sites, suppressing the energy gain of singlet formation.  The
size of these AF regions shrinks with the application of pressure, which
increases this hybridization towards its value in the absence
of disorder, ultimately yielding a revival of
superconductivity (SC).  However, the resulting phase is quite
heterogeneous\cite{seo14}, not unlike the stripe and nematic orders
which coexist with superconductivity in the cuprates.  Prior theoretical
work examined domains within the context of a mean field theory (MFT)
treatment of competing AF and SC orders\cite{seo14}.

{\it In this paper, we report QMC simulations of a disordered
bilayer Heisenberg
Hamiltonian, and characterize its physics within an exact treatment of
quantum many body fluctuations.  
Our key findings are:
(i) An appropriately defined impurity susceptibility captures both the
inhibition of AF order deep in the ordered phase,
and its sharp enhancement near the QCP;
(ii)  Quantitative determination
of the AF regions induced by impurities, and the criterion
for their coalescence into a state with long range order at
experimentally relevant temperature scales;
(iii)  Verification of the suggestion that the NMR relaxation rate is
temperature-independent at the QCP, and the demonstration that the 
local value of $1/T_1$ on an impurity site increases abruptly.}

QMC, in combination with analytic scaling arguments, has previously been
used to study the loss of magnetic order and multicritical points in
bilayers where the dilution at a given site discards simultaneously the
spins in {\it both} layers\cite{sandvik02}.  Interesting topological
considerations arise from the removal of a {\it single} spin from a
bilayer system in the singlet phase, since an unpaired spin-1/2 object
is left behind\cite{sachdev99}.  QMC has been used to study the spin
texture produced by a single impurity\cite{hoglund07} as well as the
onset of AF order in lattices of dimerized chains\cite{yasuda01}.

\vskip0.05in

\noindent

\underbar{Model and Methods:}

We consider the AF Heisenberg bilayer Hamiltonian

\begin{align}
H = \sum_{\langle i j \rangle,\alpha} 
J^{\alpha} \vec S^{\,\alpha}_{i} \cdot \vec S^{\,\alpha}_{j}
+ g \sum_{i} \vec S^{\,1}_{i} \cdot \vec S^{\,2}_{i}
\label{eq:hamiltonian}
\end{align}

Subscripts $i,j$ denote spatial sites on a square lattice, and
superscripts $\alpha=1,2$ label the two layers.  We study
the case when the intraplane exchange 
constants $J^{\alpha}=J$ are the same, and we choose $J=1$
to set the energy scale.  $g$ is the interlayer exchange.

Our treatment models disorder via site removal in one of the Heisenberg
layers.  This is an appropriate picture for the doping of Cd for Co,
which is thought to reduce sharply the hybridization of the conduction
electrons to the Ce moment.  Our motivation for studying a model of
localized spins is in part pragmatic.  Accessible system sizes for QMC
simulations of itinerant fermion systems are not sufficiently large to
encompass multiple impurities and carefully study finite size effects.
However, it is also known that many of the qualitative features of
itinerant AF models like the Hubbard Hamiltonian are reflected in their
spin counterparts\cite{foot1}, notably the successful modeling of the
Knight shift anomaly which can be captured either with descriptions in
terms of localized spins\cite{shirer12} or itinerant
electrons\cite{jiang14}.

In the absence of disorder, the position of the AF-singlet 
transition has been located by Sandvik to high 
accuracy through a finite size extrapolation of the AF order
parameter

\begin{align}
\langle m^2 \rangle = 
\left< \left(\frac{1}{N} \sum_{i} (-1)^{x_i+y_i+\alpha} \, S^{\alpha}_i \right) ^2 \right>
\label{eq:m2}
\end{align}

For the symmetric case\cite{sandvik94,wang06}, $J^1=J^2$, the
critical interlayer exchange $g_c=2.5220$.  For the `Kondo-like' lattice
where one of the intralayer $J$ is zero, $g_c=1.3888$.

As in Ref.~[\onlinecite{sandvik94}] we use the stochastic series expansion
(SSE) to obtain $\left<m^2\right>$.  SSE samples terms in a power
expansion of $e^{-\beta \hat H}$ in the partition function, using
operator loop (cluster) updates to perform the sampling
efficiently\cite{syljuasen02}.  Here we consider bilayer systems 
with $N=2 \times L\times L$ and $L$ up to $100$ sites. 

We also evaluate the NMR relaxation rate, given by the
low frequency limit of the dynamic susceptibility

\begin{align}
\frac{1}{T_1} = T \, {\rm lim}_{\omega \rightarrow 0} 
\sum_q A^2 \frac{\chi^{\prime\prime}(q,\omega)}{\omega}
\label{eq:T1}
\end{align}
where $A$ is the hyperfine coupling.
We obtain $1/T_1$ using the long imaginary-time
behavior of the spin-spin correlation function

\begin{align}
\frac{1}{T_1} = \frac{A^2}{\pi^2 T} \, \langle \, S_i(\tau=\beta/2)
S_i(\tau=0) \, \rangle
\label{eq:T1approx}
\end{align}

The regime of validity of this expression 
is discussed in Ref.~[\onlinecite{randeria92}].

\begin{figure}[]
{\centering\resizebox*{8.7cm}{!}{\includegraphics*{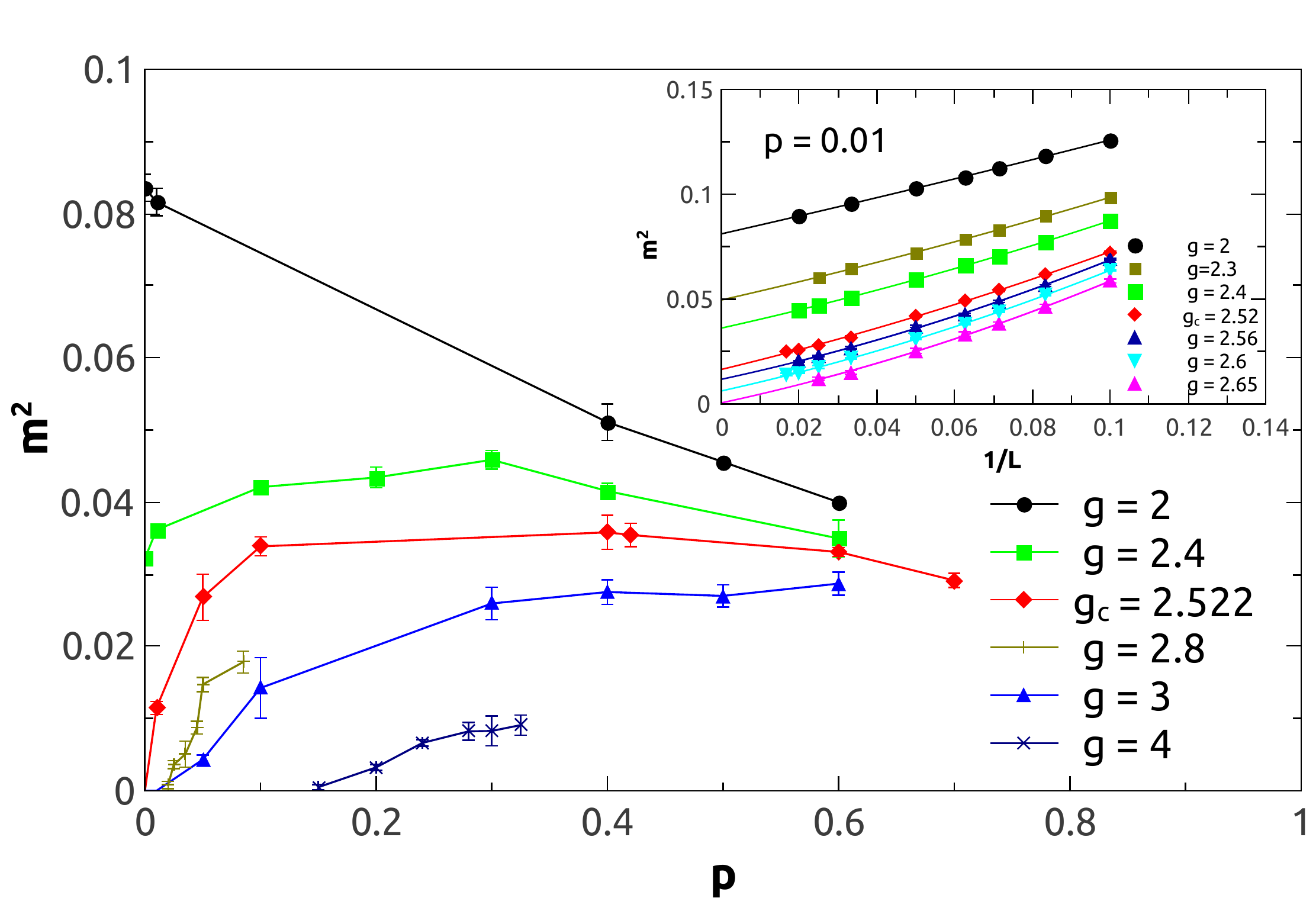}}}
\caption{The square of the staggered magnetization, $\left<m^2\right>$,
as a function of the impurity concentration, $p$, for different values
of $g$.  In the AF phase with $g=2 \ll g_c=2.522$, impurities reduce the
order.  The effect of impurities near the QCP and in the singlet phase
is discussed in the text. 
\underbar{Inset}: Finite size scaling of
$\left<m^2\right>$ for $p = 0.01$.  The position of the QCP is
increased to $g_c(p=0.01)=2.65$.
Data were averaged over 120 disorder realizations.
The inverse temperature $\beta = 80$.
} 
\label{fig1} 
\end{figure}

\vskip0.05in  \noindent

\underbar{AF Domains and Impurity Susceptibility:}
The behavior of $\left<m^2\right>$ as a function of the fraction of removed sites, $p$, is shown for a
range of $g$ in Fig.~\ref{fig1}.  
The results were averaged over $120$ dilution realizations.
As expected, impurities decrease
$\left<m^2\right>$ deep within the AF phase $(g=2)$ 
where they act to reduce the
average coordination number of the lattice and hence the tendency to
order.  Closer to the QCP, a different behavior emerges.  Impurities
begin to inhibit singlet formation by leaving unpaired moments on their
partner sites.  The AF order parameter, which had been disrupted by
singlet formation, therefore increases with $p$ for $g
\lesssim g_c$, and does so especially sharply at $g=g_c$.  For $g>g_c$
impurity concentrations $p$ which are sufficiently large can induce AF
order even though these larger interplanar couplings would 
result in singlet formation in the pure case.
The appearance of a finite $p_c$ for $g>g_c$ is discussed further below.

\begin{figure}[]
{\centering\resizebox*{8.7cm}{!}{\includegraphics*{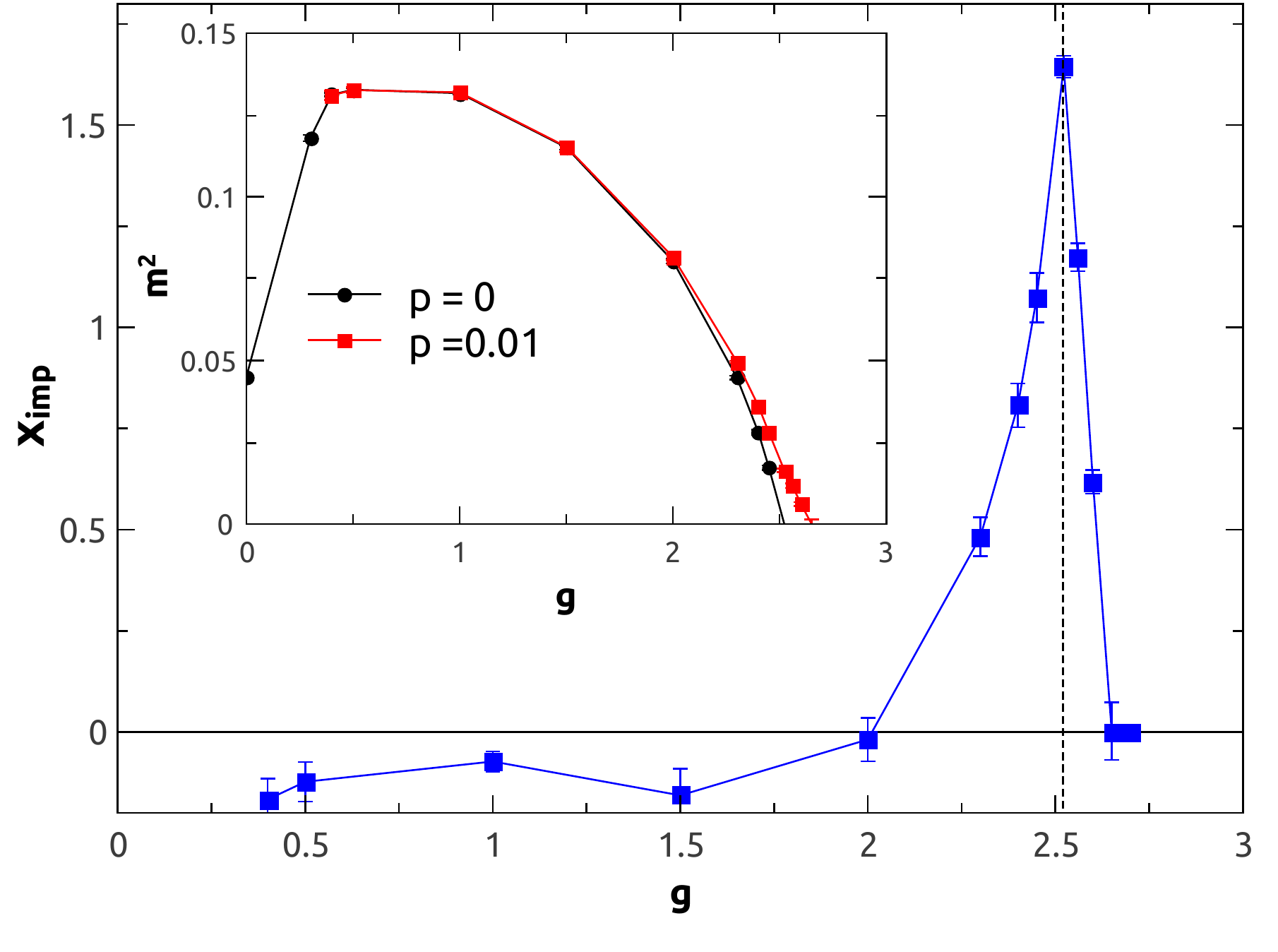}}}
\caption{The impurity susceptibility, 
$\chi_{\rm imp}$, as function of $g$ is sharply peaked at $g_c$
(vertical dashed line): impurities induce AF order.  
Away from $g_c$, $\chi_{\rm imp} < 0$:
impurities reduce the AF order parameter.
\underbar{Inset}: shows the $g$ dependence of $\left<m^2\right>$ for $p =
0.01$ (square) and clean system (circles).  Both the shift in $g_c$ and
the large effect of impurities at the QCP are evident.
Data for $\left<m^2\right>$ have been extrapolated to $L=\infty$.
The inverse temperature $\beta=80$.
} 
\label{fig4} 
\end{figure}

The effect of impurities on the AF order parameter can be characterized
by an `impurity susceptibility'
\begin{equation}
\chi_{\rm imp}=\frac{d \left<m^2\right>}{dp} \Bigg| _{p=0},
\label{eq:Chi}
\end{equation}
as shown in Fig.~\ref{fig4}.  $\chi_{\rm imp}$ 
has a sharp peak at $g_c$.  That is, the
effect of the impurities is especially large close to the QCP where the
system in delicately poised between two phases.  Farther away
from the QCP in the AF, $g \lesssim 2$, the impurity
susceptibility is negative, as can be inferred from 
related data for the 2D Heisenberg
model with site-dilution \cite{sandvik02_2}.

For $g > g_c$, impurities induce AF order in an otherwise singlet phase
\cite{sandvik97}.  We can estimate the critical impurity concentration
as follows:  Prior to the establishment of order, the coupling between
two regions centered at sites $i$ and $j$ will oscillate in phase, with
an amplitude which decays exponentially 
\cite{sigrist96,hass01},
$J_{\rm eff} \approx J
(-1)^{-|i - j +1|} {\rm exp}(-\left<l\right> /\xi)$.
Here $\left<l\right>$ is the mean impurity
separation and $\xi$ is the correlation length in the clean system.  For
the 2D system considered here, $\left<l\right> = 1/\sqrt{p}$.  Assuming
that the AF order will set in when the average distance between the
impurities is on the same scale as
$\xi$, yields the condition $\xi \sqrt{p_c} \approx 1$.  Assuming the
system is dilute, we can compute $\xi$ by embedding a single impurity in
the lattice and evaluating the decay of the spin-spin correlation in its
vicinity;  See Supplemental Materials.  
Fig.~\ref{fig2}(a) shows the resulting $\xi$.  The center panel validates
the picture that the critical concentration of impurities to induce AF
order occurs when $\langle l \rangle = 1/\sqrt{p} \propto \xi$.

\begin{figure}[]
{\centering\resizebox*{8.7cm}{!}{\includegraphics*{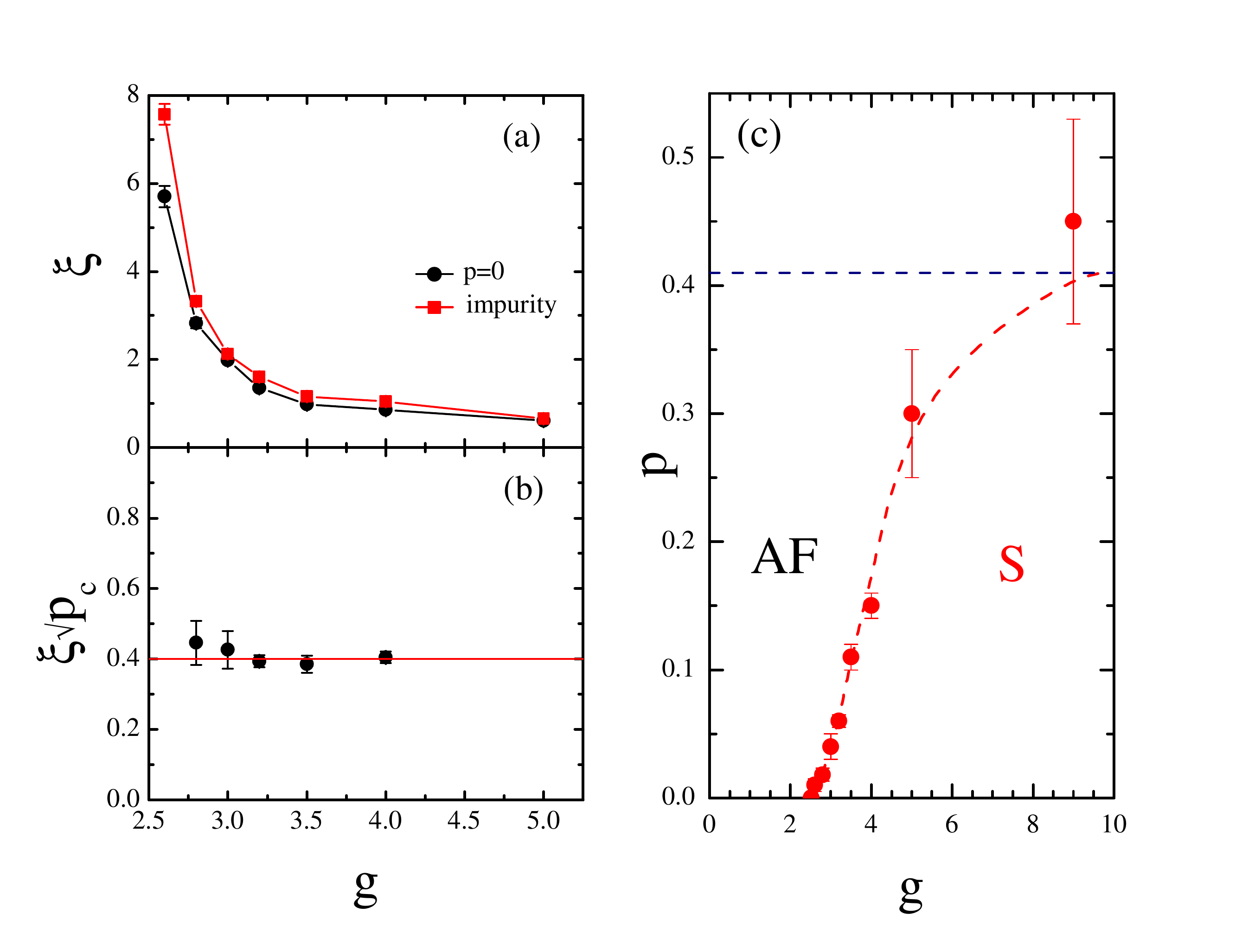}}}
\caption{(a) Correlation length  $\xi$ as function of $g$. 
Data are shown both around a
single impurity (square) and for the clean system  (circles). 
(b) $\xi \sqrt{p_c}$ as a function of $g$ is roughly constant,
consistent with a qualitative picture where AF order occurs when the
mean impurity separation $\langle l \rangle$ is proportional to $\xi$.  
See text.
Data for inverse temperatures $\beta=40,80$ were compared
to ensure convergence to the ground state.
$L$ up to $100$ was used to calculate $\xi$.
(c) The AF order parameter at fixed $\beta=80$ exhibits a sharp
crossover indicating the position of the enhanced range of AF order
created by the spin-1/2 impurities left behind on a fraction $p$
of sites.
} 
\label{fig2} 
\end{figure}

There are several subtleties to this argument.  At $T=0$, it has been 
argued that an
exponentially small interaction between impurities can induce
order\cite{laflorencie04}. (This occurs despite the fact
that some impurity pairs, which are
sufficiently close spatially, lock into singlets\cite{roscilde06}).  This
suggests $p_c=0$ throughout the singlet phase- an arbitrarily small
number of impurities will order despite the rapid decay of their
coupling.  The effect of these very small couplings is, however, seen 
only at extremely low $T$, 
a fact that is reflected in SSE
simulations\cite{laflorencie04} by the need to study inverse
temperatures $\beta$ as large as $\beta \sim 10^4-10^5$ (except very
close to the QCP where $\xi$ diverges).  In contrast, $\beta$ which is
2-3 orders of magnitude smaller is sufficient to reach the ground state
on lattices of $L\sim 60$ studied here.  

The ordering temperatures observed in Cd-doped CeCoIn5
are about 2-5 K, and the c-f
coupling is reported to be around 49 meV, so that
$T_c \sim 10^{-2} J$.
Thus a more refined
interpretation of Fig.~\ref{fig2}(b) is that,
although AF likely exists at infinitesmal $p_c$ strictly
at $T=0$, panel (b)
gives the effective critical impurity
concentration to induce AF in the singlet phase
at experimental temperature scales\cite{curro07}.
Fig.~\ref{fig2}(c) shows the position of the sharp cross-over which
occurs in the AF order parameter at these scales.

\vskip0.05in  \noindent

\underbar{Universal Behavior of the NMR relaxation rate:}
The NMR spin relaxation rate, $1/T_1$ provides a crucial experimental
window into doped heavy fermion materials where secondary peaks in the
spectra, and broadening of the main spectral line implicate the presence
of inhomogeneous environments\cite{seo14}.  In this section we provide a
quantitative description of the effect of impurities on $1/T_1$ and
demonstrate that these provide a crisp signature at the QCP,
shown to higher resolution in the inset.

\begin{figure}[]
{\centering\resizebox*{9cm}{!}{\includegraphics*{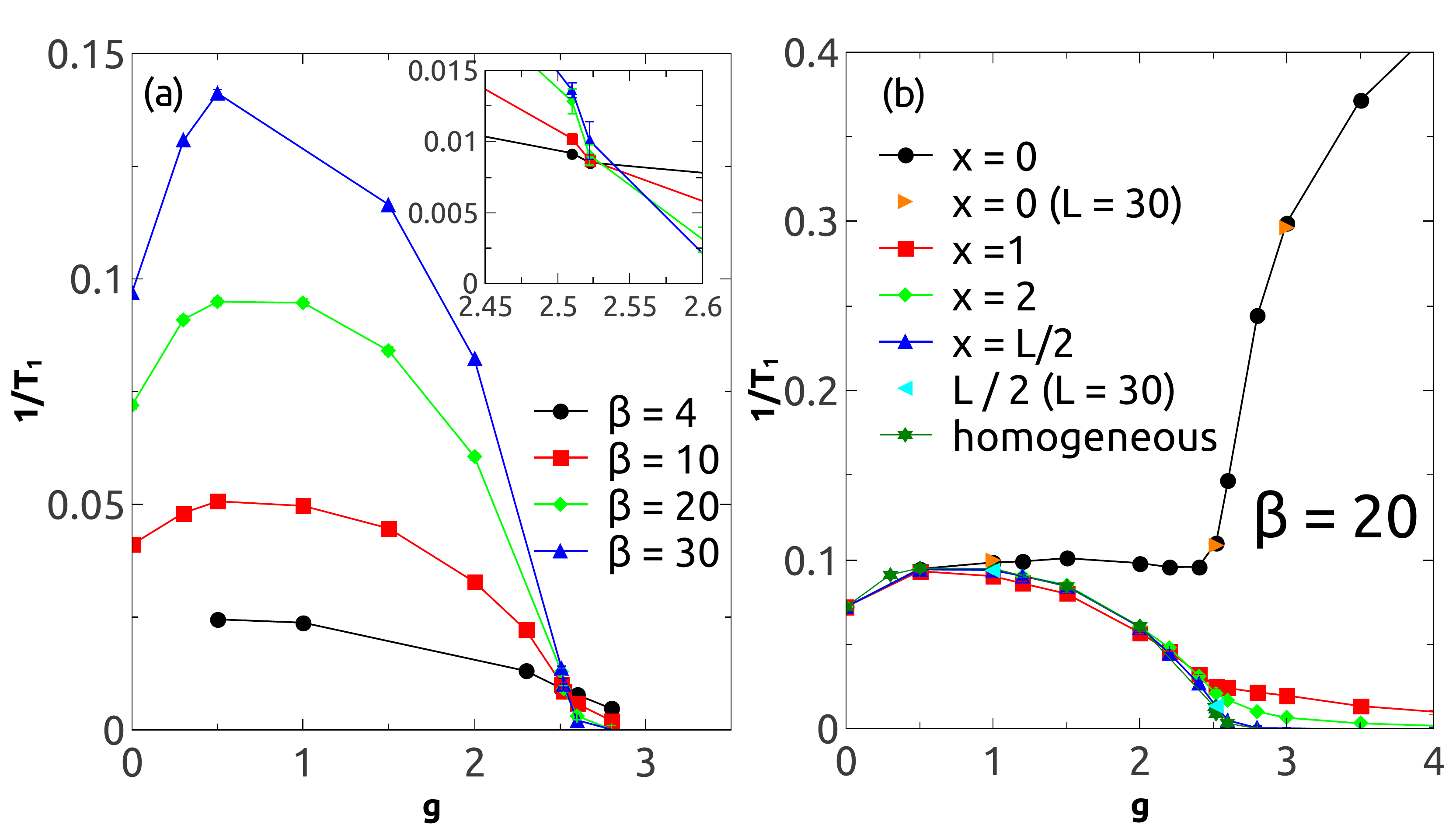}}}
\caption{(a) $1/T_1$ as a function of $g$ for different values of $\beta$. 
The linear size $L = 50, 60$.
\underbar{Inset}: Blow-up of the crossing point.
$g < g_c$, $g = g_c$ and $g > g_c$.
(b) $g$ dependence of $1/T_1$ for a system with a single removed spin.
$x= 0, 1, 2, L/2$ are different horizontal distances from the 
impurity. See text.   For the impurity system the linear lattice size $L=20, 30$.}
\label{fig5} 
\end{figure}

The main panel of
Fig.~\ref{fig5}a shows the evolution of $1/T_1$ with interlayer coupling
at different fixed temperatures.  $1/T_1$ follows the same trend as
the AF order parameter $\left<m^2\right>$, i.e.~it initially rises as
the two planes are coupled, has a maximum for $g \approx 0.5$, and then
decreases to small values at the QCP.  
The $T$ independence of $1/T_1$ \cite{sachdev93,qimiao01}, see the Supplementa material, is emphasized by the common crossing point at $g_c \sim 2.52$, see inset of Fig.~\ref{fig5}a.

The behavior of $1/T_1$ in the presence of disorder
is shown in Fig.~\ref{fig5}b.
We consider the most simple case where a spin is removed from one layer,
and compute the relaxation rate of spins in the pure layer
as a function of distance $x$
on a horizontal line from the removed site. $x=0$ in the figure thus
corresponds to the removed spin's immediate partner, while
$x=1,2$ are near and next-near neighbors, and, finally, 
at $x=L/2$, far away from the impurity.
For $x=0$ the partner spin shows a sharp signature of the QCP.
Above $g=g_c$, when all the other spins are locked in singlets,
the free spin-1/2 left behind by the spin removal has a greatly enhanced
$1/T_1$.  Meanwhile, the relaxation rate is small for all
other sites.  
For $g=0$ the spins in the undiluted plane are decoupled from
the second layer, and hence will all share a common value of $1/T_1$
regardless of impurites placed there.  
Figure \ref{fig5}b indicates that this independent plane
behavior approximately extends out to $g \lesssim g_c$ for $x \ge 1$.  
The curve for $x=0$ breaks away for $g \ge 1 $, 
and, as noted earlier, has a very sharp increase at the QCP.
The comparison with the curve of the homogeneous system shows, as is observed experimentally \cite{curro07}, that the $1/T_1$ of the farthest spin is unaffected by the impurity.
In the Supplemental Material 
further details on the spatial distribution of 
$1/T_1$ are provided.

\begin{figure}[]
{\centering\resizebox*{8.7cm}{!}{\includegraphics*{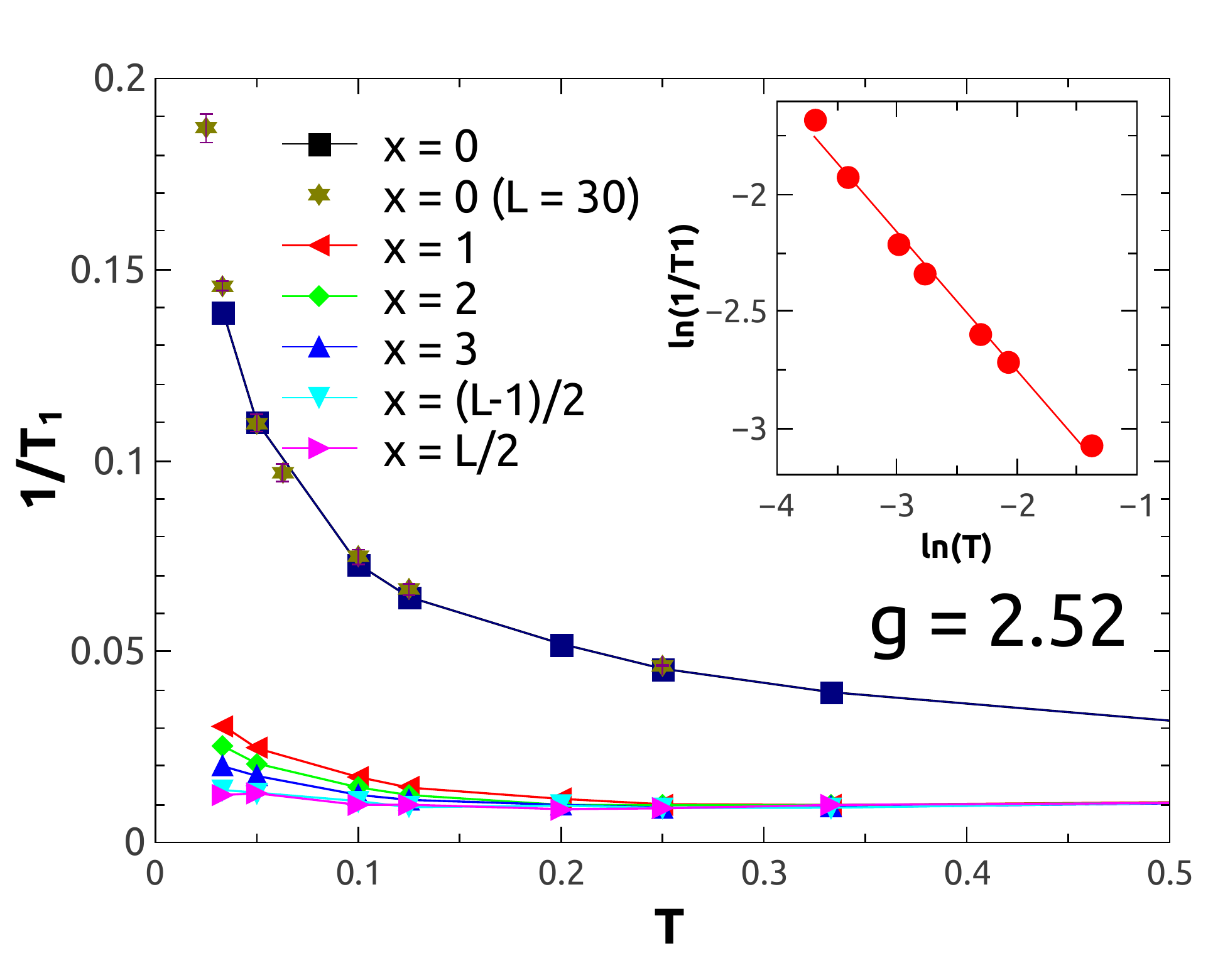}}}
\caption{$T$ dependence of $1/T_1$ for separations $x$ from the
impurity.
See text.  \underbar{Inset}: Determination of the
$\eta^\prime$ exponent.  Linear lattices sizes were $L=20,30$, somewhat
smaller than in previous figures because of the necessity to compute the
imaginary time dependent correlation functions.}
\label{fig6} 
\end{figure}

We conclude our discussion of the relaxation rate by computing the
temperature dependence of $1/T_1$ at the QCP, $g=2.52$,
for this same collection of sites.
As emphasized by Fig.~\ref{fig6}, $1/T_1$ is only weakly temperature
dependent away from the impurity site.
For the spin left behind at $x=0$, $1/T_1$ increases substantially
as $T$ is lowered.  The growth can be described by a power law.
(See inset.)
Sachdev \textit{et.~al.} have argued
\cite{sachdev99} that the imaginary-time autocorrelation
function of an impurity at the QCP follows the scaling
form, $S_i(\tau)S_i(0) \sim  \tau^{\eta^\prime}$, which implies,
(Eq.~\ref{eq:T1approx}), that
$1/T_1 \sim T^{(\eta^{\prime} - 1)}$.
Here we obtain $\eta^{\prime} = 0.41 \pm 0.03$, in agreement with 
Ref.~[\onlinecite{hoglund07}].

\begin{figure}[]
{\centering\resizebox*{8.7cm}{!}{\includegraphics*{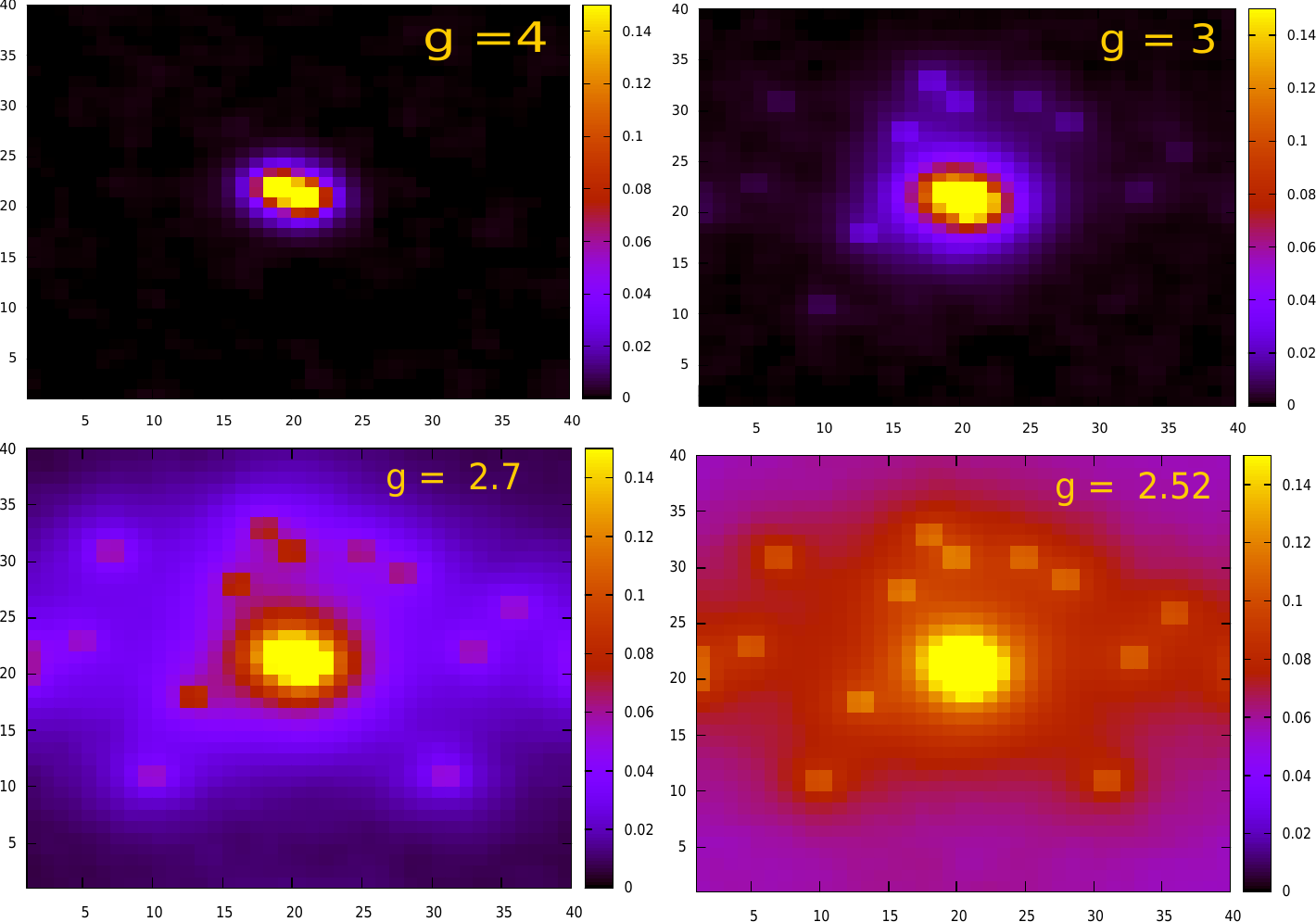}}}
\caption{Correlation of the spin at an impurity site, chosen
to be at the center of the lattice, with spins at other sites.  Deep in the
singlet phase $g=4$, top left, the ordered region is confined to
the immediate vicinity of the site.  As $g$ decreases towards the QCP,
this puddle first expands, top right, followed by the emergence of large
values at the locations of the unpaired spins at other impurity sites,
bottom left and right.
\label{figX} 
}
\end{figure}

Finally, in Fig.~\ref{figX}, we study the AF puddles which are believed
to form around Cd sites, and their interactions.  For $g=4$ only spins
in the close vicinity of an impurity at the lattice center are
correlated with it.  When $g=3$ the locations of the other impurities
become evident.  Correlations starting to become substantial over
the whole lattice at $g\sim 2.7$.  For this density of impurities,
$p=0.01$ (16 impurities on a 40x40 lattice) finite size scaling showed
that $p_c \sim 2.52$.  (See inset to Fig.~\ref{fig1}.) The enhanced
values of the spin correlations at the clean system QCP (bottom
right panel) are consistent with the establishment of a non-zero value
for the order parameter, $m^2 \sim 0.02$ in Fig.~\ref{fig1}.  
Results for the puddle formed by a single impurity are in the
Supplementary materials.

\underbar{Conclusions and Outlook:}
Exploration of the effect of randomness and dilution on magnetic
and superconducting order is a crucial step to understanding 
disordered, strongly interacting quantum systems such as heavy fermion and
cuprate materials, particularly near quantum critical points.
Impurities reduce order, but can also nucleate ordered domains
which, when sufficiently close to one another, can coalesce to create long range 
order \cite{millis01,zhu02,andersen07}.
In this paper we have brought together exact QMC 
calculations of the effect of impurities on
spin correlations/domains and the nuclear magnetic resonance relaxation
rate as a system is tuned through a QCP.

Our key conclusions are that:
(i)  the impurity susceptibility, defined as the response of the
AF order parameter to the removal of a small number of spins,
exhibits a sharp peak at the QCP, so that low disorder concentrations
readily lead to long range order;
(ii)  The critical concentration $p_c$ for randomness to induce long
range AF order in the singlet phase, at moderate $\beta$,
is well described by $\xi \, \sqrt{p_c} \sim 0.4$, where $\xi$
is the spin correlation length
at $g > g_c$;
and (iii)  
Verification of the suggestion that the NMR relaxation rate is
temperature-independent at the QCP, and that an abrupt increase in the
local value of $1/T_1$ on an impurity site provides 
a clear signature of the QCP.

Our work has focussed on localized Heisenberg spins.
Analogous studies within itinerant electron Hamiltonians
like the periodic Anderson model, are 
underway\cite{benali16}.

\vskip0.15in

\underbar{Acknowledgements:}
We would like to thank Rajiv Singh and Eric Andrade
for very useful discussions.
T.P. and R.R.S. acknowledges support from CNPq and CAPES,
FAPERJ, and the INCT on Quantum Information. 
R.T.S. and N.C. acknowledge support from the
NNSA grant DE-NA0002908.
T.M., T.P. and R.T.S. acknowledge
funding from Science Without Borders, Brazil.
G.B. acknowledges support from Institut Universitaire de France, 
MajuLab, and Centre for Quantum Technologies.
R.T.S.~and G.B.~thank T. Fratellis.

\renewcommand{\thefigure}{S\arabic{figure}}

\setcounter{figure}{0}

\begin{figure}[h!]
{\centering\resizebox*{8.7cm}{!}{\includegraphics*{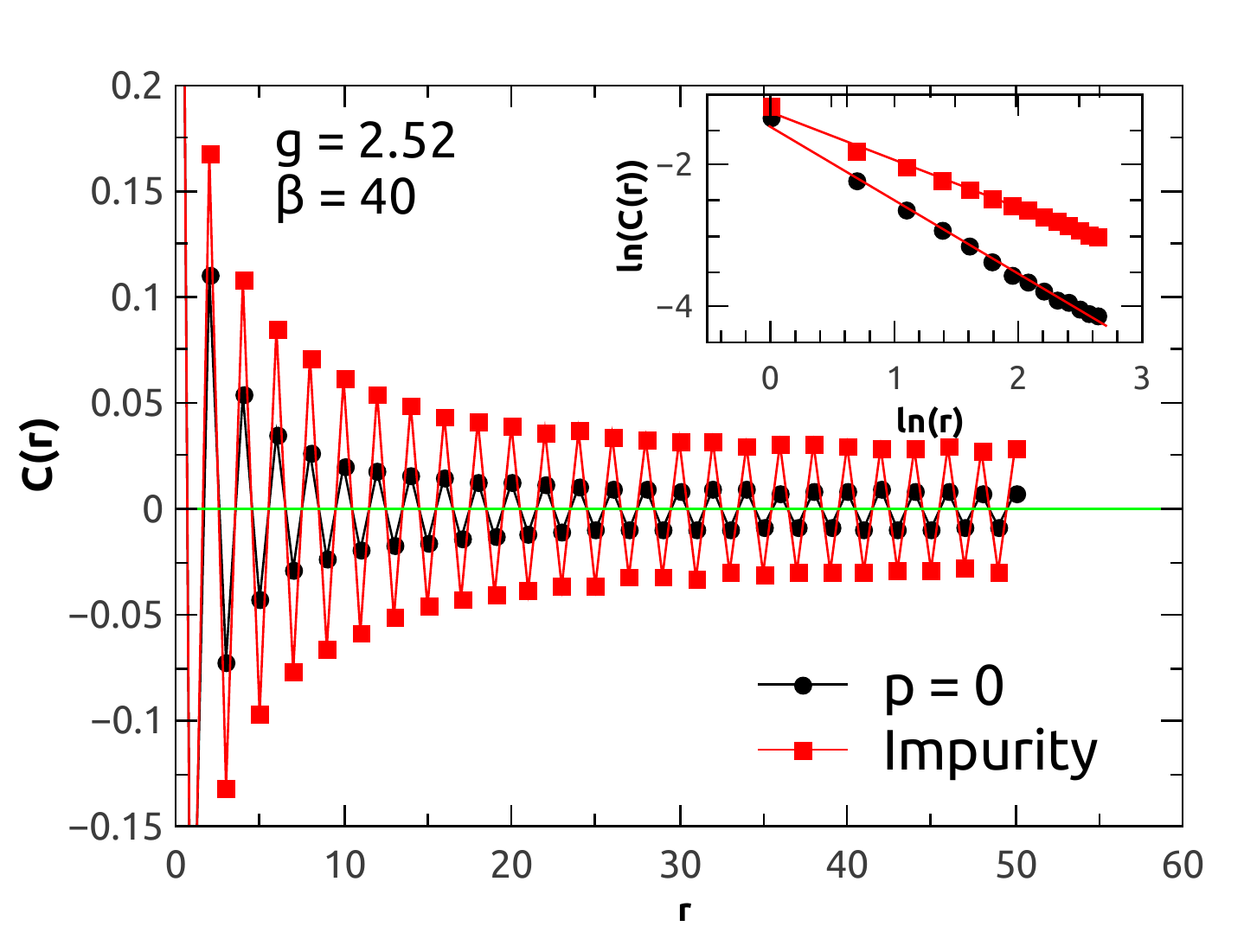}}}
\caption{Staggered spin correlation function, $C(r)$, at $g = g_c$ for a
single impurity (square) and clean system (circle).  \underbar{Inset}:
Power law behavior of $C(r)$.  
A stronger AF cloud is produced about
the unpaired impurity moment.
} 
\label{figS1} 
\end{figure}

\underbar{Supplementary Material:}
At the QCP, Fig.~(1) of the main text showed 
that $\left<m^2\right>$ is 
significantly different from zero even for a small fraction of impurities,
$p = 0.01$.  
In Fig.~\ref{figS1} we provide further details
by showing the staggered correlation
function, $C(r) = (-1)^{(i+r)} S_i^1  S_{i+r}^1$, of a single impurity
at $g = g_c$.  Here $S_i^1$ is the spin in layer 1
left without a partner by the
dilution at site $i$.  $C(r)$ is enhanced around such a 
``lonely site", in comparison with $C(r)$ in the
clean system (also shown).  

The log-log plot in the inset shows the decay is power law.
In the clean system,
$C(r) \sim 1/r^{1+\eta}$, where $\eta = 0.03$ is the 3D
Heisenberg model critical exponent.  The impurity $C(r)$ also follows a
power law decay, but with an exponent $\alpha \approx 0.67 \pm 0.02 < 1 + \eta$.
For $g>g_c$ the singlet gap opens and $C(r)$ decays exponentially,
as discussed in the main text.

\begin{figure}[h!]
{\centering\resizebox*{8.7cm}{!}{\includegraphics*{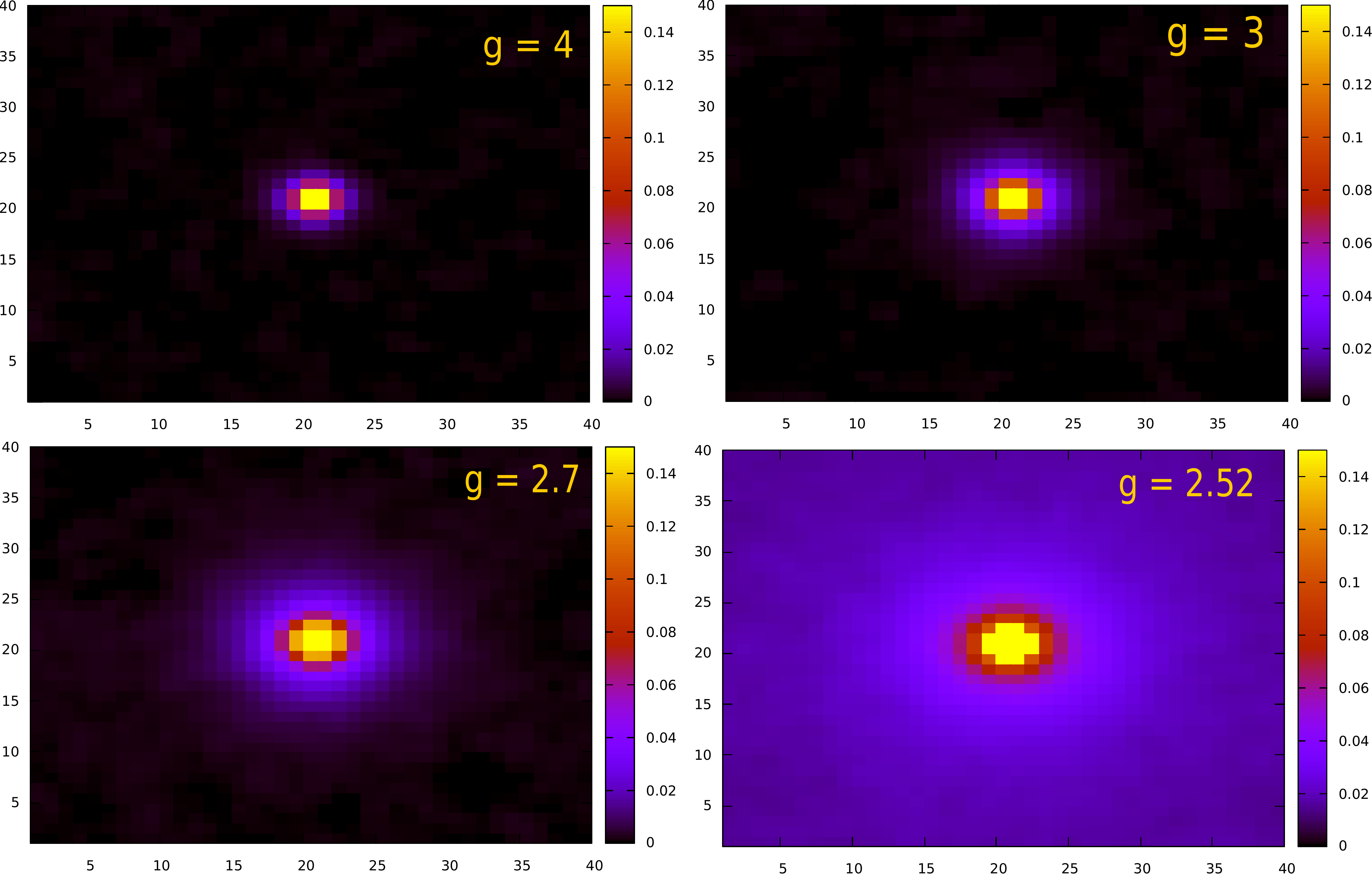}}}
\caption{Color contour plot of the AFM cloud around a {\it single} impurity.  Deep in the singlet phase, $g=4$ (upper left) spin
correlations extend for only a few lattice spacings.  This distance
grows as $g$ approaches $g_c$ from above.  
\label{figXS} 
}
\end{figure}

Fig.~(6) of the main text showed 
the AF clouds around a collection of impurities for various depths into
the singlet phase.
Here Fig.~\ref{figXS} illustrates the cloud produced by a single
impurity.  
The line graph of Fig.~\ref{figS1} corresponds to a horizontal cut of the 
bottom right panel of Fig.~\ref{figXS}.

\begin{figure}[h]
{\centering\resizebox*{8.7cm}{!}{\includegraphics*{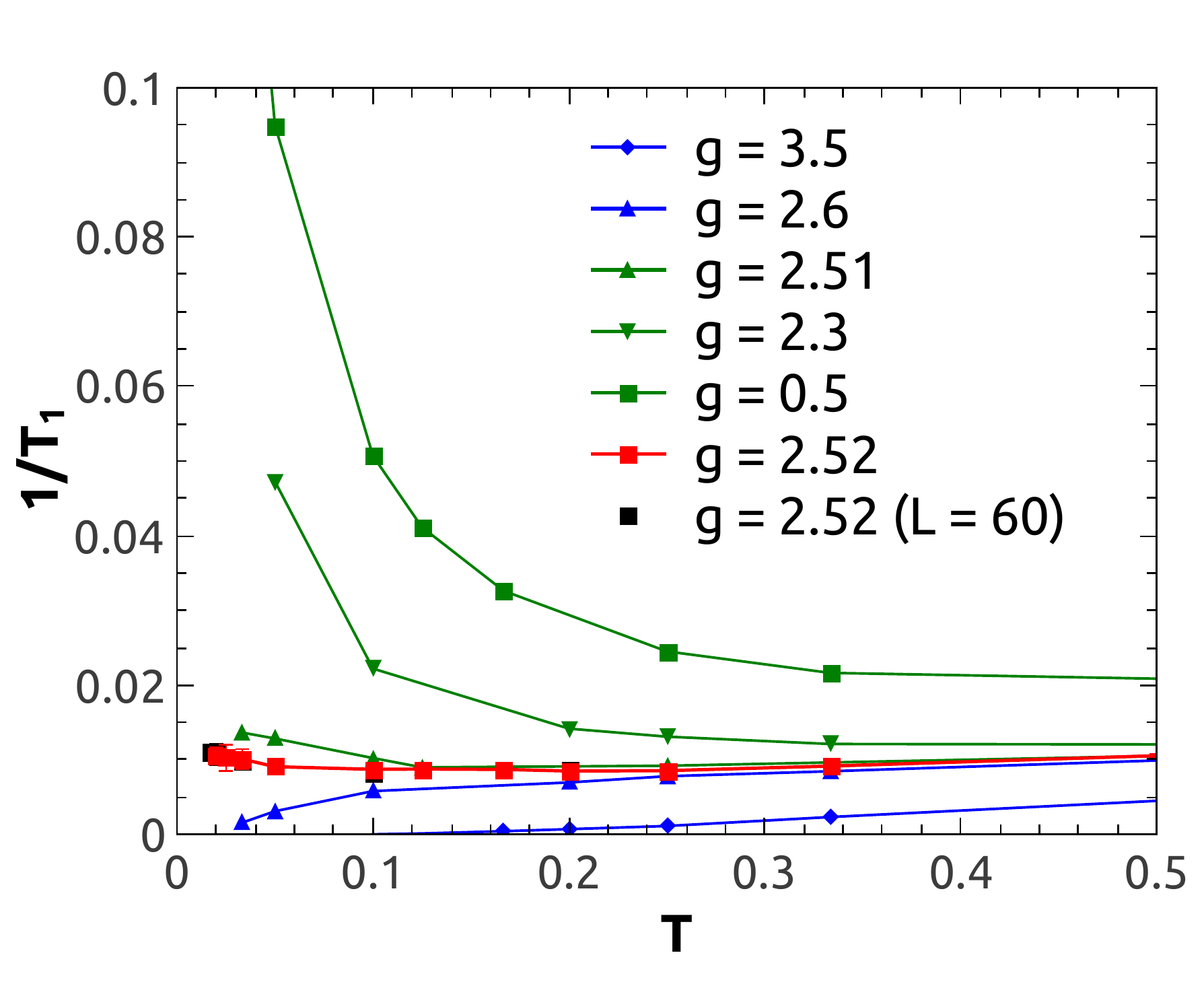}}}
\caption{$1/T_1$ as a function of $T$ for different 
values of $g$ in the homogeneous system.
Linear size of the system $50, 60$.
At the QCP, $g=2.52$, the relaxation rate is nearly constant.
} 
\label{T1vsT} 
\end{figure}

In Fig.~(\ref{T1vsT}),
we re-display the data of Fig.~(5),
showing now $1/T_1$ as a function of temperature $T$. 
In the AF phase ($g < g_c$), $1/T_1$
increases for low $T$.  On the other hand, due to the presence of the
spin gap, $1/T_1$ goes to zero in the singlet phase ($g > g_c$).  At the
QCP, $g = g_c$, $1/T_1$ is almost constant for the range of temperature
considered \cite{sachdev93,qimiao01}.
This is an alternate view of the crossing of the curves of
$1/T_1$ versus $g$ shown in Fig.~(4)a.

\begin{thebibliography}{100}







\bibitem{Takagi89}
``Superconductor-to-nonsuperconductor transition in
(La$_{1-x}$Sr$_x$)$_2$CuO$_4$ as
investigated by transport and magnetic measurements,"
H. Takagi, T. Ido, S. Ishibashi, M. Uota, S. Uchida, and Y. Tokura,
Phys. Rev. B40, 2254 (1989).


\bibitem{Uchida91}
``Optical spectra of La$_{2-x}$Sr$_x$CuO$_4$: Effect of carrier doping on the
electronic structure of the CuO2 plane,"
S. Uchida, T. Ido, H. Takagi, T. Arima, Y. Tokura, and S. Tajima,
Phys. Rev. B43, 7942 (1991).


\bibitem{xiao88}
``High-temperature superconductivity in tetragonal perovskite structures:
Is oxygen-vacancy order important?",
Gang Xiao, M. Z. Cieplak, A. Gavrin, F. H. Streitz, A. Bakhshai, and C.
L. Chien,
Phys. Rev. Lett. 60, 1446 (1988).


\bibitem{keimer92}
``N\'eel transition and sublattice magnetization of pure and doped
La$_2$CuO$_4$",
B. Keimer, A. Aharony, A. Auerbach, R. J. Birgeneau, A. Cassanho, Y.
Endoh, R. W. Erwin, M. A. Kastner, and G. Shirane,
Phys. Rev. B45, 7430 (1992).



\bibitem{mahajan94}
``$^{89}$NMR probe of Zn induced local moments in 
YBa$_2$(Cu$_{1-y}$Zn$_y$)$_3$O$_{6+x}$",
A. V. Mahajan, H. Alloul, G. Collin, and J. F. Marucco,
Phys. Rev. Lett. 72, 3100 (1994).


\bibitem{ido91}
``Optical study of the La$_{2-x}$Sr$_x$NiO$_4$ system: 
Effect of hole doping on the
electronic structure of the NiO$_2$ plane,"
T. Ido, K. Magoshi, H. Eisaki, and S. Uchida,
Phys. Rev. B44, 12094(R) (1991).



\bibitem{reinert95}
``Electron and hole doping in NiO,"
F. Reinert, P. Steiner, S. H\"ufner, H. Schmitt, J. Fink, M. Knupfer,
P. Sandl, and E. Bertel, Z. Phys. B97, 83 (1995).


\bibitem{martin97}
``Spin-Peierls and antiferromagnetic phases in 
Cu$_{1-x}$Zn$_x$GeO$_3$: A
neutron-scattering study",
Michael C. Martin, M. Hase, K. Hirota, G. Shirane, Y. Sasago, N. Koide,
and K. Uchinokura,
Phys. Rev. B56, 3173 (1997).


\bibitem{azuma97}
``Switching of the gapped singlet spin-liquid state to an
antiferromagnetically ordered state in Sr(Cu$_{1-x}$Zn$_x$)$_2$O$_3$,"
M. Azuma, Y. Fujishiro, M. Takano, M. Nohara, and H. Takagi,
Phys. Rev. B55, R8658 (1997).


\bibitem{ulmke99}
``Disorder and Impurities in Hubbard Antiferromagnets",
M.~Ulmke, P.~J.~H.~Denteneer, V.~Janis, R.~T.~Scalettar, A.~Singh,
D.~Vollhardt, and G.~T.~Zimanyi,
Adv. Sol. St. Phys. 38, 369 (1999).


\bibitem{hoglund04}
``Impurity effects at finite temperature in the two-dimensional
$S=1/2$ Heisenberg antiferromagnet,"
K.H. Hoglund and A.W. Sandvik,
Phys. Rev. B70, 024406 (2004).


\bibitem{pham06}
``Reversible tuning of the heavy-fermion ground state in CeCoIn$_5$", 
L.D. Pham, T. Park, S. Maquilon, J.D. Thompson, and Z. Fisk,
Phys.  Rev. Lett. 97, 056404 (2006). 


\bibitem{seo14} 
``Disorder in quantum critical superconductors," S. Seo,	Xin Lu,	J-X.
Zhu,	R.R. Urbano,	N. Curro,	E.D. Bauer,	V.A. Sidorov,	L.D.
Pham,	Tuson Park,	Z. Fisk	and J.D. Thompson,
Nature Physics 10, 120 (2014).



\bibitem{foot1}
In fact, bilayer Hubbard and Heisenberg models can even track each other
semi-quantitatively.  For example, the Hubbard QCP $(t_\perp/t)_c \sim
1.6$ is rather close to that obtained from the Heisenberg limit
$(J_\perp/J)_c =2.522$ assuming $J=4t^2/U$.  See `Magnetic and Pairing
Correlations in Coupled Hubbard Planes." R.T.~Scalettar, J.W.~Cannon,
D.J.~Scalapino, and R.L.~Sugar, Phys.~Rev.~B50, 13419 (1994).



\bibitem{shirer12}
``Long range order and two-fluid behavior in heavy
electron materials,"
K.R. Shirer, A.C. Shockley, A.P. Dioguardi, J. Crocker,
C.-H. Lin, N. apRoberts-Warren, D. M. Nisson, P.  Klavins,
J.C. Cooley, Y.-F. Yang, and N. J. Curro,
Proc.~Nat.~Acad.~Sci.~109, 18249 (2012).



\bibitem{jiang14}
``Universal Knight shift anomaly in the Periodic Anderson model",
M.~Jiang, N.J.~Curro, and R.T.~Scalettar,
Phys. Rev. B90, 241109 (2014).



\bibitem{held00}
``Similarities between  the Hubbard and Periodic
Anderson Models at  Finite Temperatures",
K.~Held, C.~Huscroft, R.T.~Scalettar, and A.K.~McMahan,
Phys.~Rev.~Lett.~85, 373 (2000).


\bibitem{sandvik02}
``Multicritical point in a Diluted Bilayer Heisenberg Quantum
Antiferromagnet,"
A.W. Sandvik, 
Phys. Rev. Lett. 89, 177201 (2002).



\bibitem{sachdev99}
``Quantum impurity in a nearly-critical two dimensional
antiferromagnet,"
S. Sachdev, C. Buragohain, and M. Vojta, Science 286, 2479 (1999).



\bibitem{hoglund07}
``Impurity Induced Spin Texture in Quantum Critical 2D
Antiferromagnets,"
K.H. H\"oglund, A.W. Sandvik, and S. Sachdev,
Phys. Rev. Lett. 98, 087203 (2007).


\bibitem{yasuda01}
``Site-dilution-induced antiferromagnetic long-range order in a
two-dimensional spin-gapped Heisenberg antiferromagnet,"
C. Yasuda, S. Todo, M. Matsumoto, and H. Takayama,
Phys. Rev. B64, 092405 (2001).

\bibitem{sandvik94}
``Order-disorder transition in a two-layer quantum antiferromagnet,"
A.W. Sandvik and D.J. Scalapino,
Phys. Rev. Lett. 72, 2777 (1994).





\bibitem{wang06} 
``High-precision finite-size scaling analysis of the quantum critical
point of $S=1/2$ Heisenberg antiferromagnetic bilayers",
L. Wang, K.S.D. Beach, and A.W. Sandvik,
Phys. Rev. B73, 014421 (2006).





\bibitem{syljuasen02}
``Quantum Monte Carlo with Directed
Loops", O.F. Syljuasen and A. W. Sandvik, 
Phys. Rev. E66, 046701 (2002). 



\bibitem{randeria92} 
``Pairing and Spin Gap in the Normal State of Short Coherence
Length Superconductors",
M.~Randeria, N.~Trivedi, A.~Moreo, and R.T.~Scalettar,
Phys.~Rev.~Lett. 69, 2001 (1992).


\bibitem{sandvik97}
``Nonmagnetic impurities in spin-gapped and gapless
Heisenberg antiferromagnets", 
A. W. Sandvik, E. Dagotto, and D. J. Scalapino, 
Phys. Rev. B56, 11701 (1997).



\bibitem{sigrist96}
``Low-Temperature Properties of the Randomly Depleted
Heisenberg Ladder'',
M. Sigrist and A. Furusaki, 
J. Phys. Soc. Jpn. 65, 2385 (1996)

\bibitem{hass01}
``Order by Disorder from Nonmagnetic Impurities in
a Two-Dimensional Quantum Spin Liquid'',
Stefan Wessel, B. Normand, M. Sigrist, and S. Haas,
Phys. Rev. Lett. 86, 1086 (2001).







\bibitem{mendels09}
``Impurity-Induced Magnetic Order in Low-Dimensional
Spin-Gapped Materials'', 
J. Bobroff, N. Laflorencie, L. K. Alexander, A.V. Mahajan, B.
Koteswararao, and P.  Mendels, Phys. Rev. Lett. 103, 047201 (2009).







\bibitem{laflorencie04}
``Magnetic ordering in a doped frustrated spin-Peierls system,"
N. Laflorencie, D. Poilblanc, and A.W. Sandvik,
Phys. Rev. B69, 212412 (2004).





\bibitem{roscilde06}
``Field-induced quantum-disordered phases in $S=1/2$ weakly coupled
dimer systems with site dilution,"
T. Roscilde,
Phys. Rev. B74, 144418 (2006).


\bibitem{millis01}
``Local defect in metallic quantum critical systems,"
A.J. Millis, D.K. Morr, and J. Schmalian, 
Phys. Rev. Lett. 87, 167202 (2001).


\bibitem{zhu02}
``Spin and charge order around vortices and impurities in high-$T_c$
superconductors," 
J-X. Zhu, I. Martin, and  A.R. Bishop, 
Phys. Rev. Lett.  89, 067003 (2002).


\bibitem{andersen07}
``Disorder-induced static antiferromagnetism in cuprate
superconductors,"
B.M. Andersen, P.J. Hirschfeld, A.P. Kampf, and M. Schmid,
Phys. Rev. Lett. 99, 147002 (2007). 



\bibitem{benali16}
``Impurity-Induced Antiferromagnetic Domains in the Periodic Anderson
Model," A. Benali, Z. Bai, N.J. Curro, and R.T. Scalettar,
arXiv:1604.02735.


\bibitem{sachdev93}
``Universal Magnetic Properties of La$_{2-\delta}$Sr$_{\delta}$CuO$_4$ 
at Intermediate Temperatures``
A.V. Chubukov and S. Sachdev,
Phys. Rev. Lett. 71, 169 (1993).



\bibitem{qimiao01}
``Locally critical quantum phase transitions in strongly correlated metals``
Q. Si, S. Rabello, K. Ingersent, and J.L. Smith,
Nature 413, 804 (2001).



\bibitem{sandvik02_2}
``Classical percolation transition in the diluted two-dimensional $S=1/2$ Heisenberg antiferromagnet``
A.W. Sandvik,
Phys. Rev. B 66, 024418 (2002).


\bibitem{curro07}
``Interecting Antiferromagnetism Droplets in Quantum Critical CeCoIn$_5$``
R.R. Urbano \textit{el. al}, 
Phys. Rev. Lett. 99, 146402 (2007).


\end{thebibliography}
\end{document}